\documentclass[twocolumn]{aastex62}
\graphicspath{{./}{figures/}}

\received{2017 August 7}
\accepted{2017 August 16}
\submitjournal{ApJL}

\shorttitle{Inversion angle of phase-polarization curve of Phaethon}
\shortauthors{Shinnaka et al.}

\begin{document}

\title{Inversion angle of phase-polarization curve of near-Earth asteroid (3200) Phaethon}

\correspondingauthor{Yoshiharu Shinnaka}
\email{yoshiharu.shinnaka@cc.kyoto-su.ac.jp, yoshiharu.shinnaka@nao.ac.jp}

\author[0000-0003-4490-9307]{Yoshiharu Shinnaka}
\affiliation{Laboratory of Infrared High-resolution Spectroscopy (LiH), Koyama Astronomical Observatory, Kyoto Sangyo University, Motoyama, Kamigamo, Kita-ku, Kyoto 603-8555, Japan}
\affiliation{National Astronomical Observatory of Japan, 2-21-1 Osawa, Mitaka, Tokyo 181-8588, Japan}

\author[0000-0001-5903-7391]{Toshihiro Kasuga}
\affiliation{National Astronomical Observatory of Japan, 2-21-1 Osawa, Mitaka, Tokyo 181-8588, Japan}
\affiliation{Department of Physics, Kyoto Sangyo University, Motoyama, Kamigamo, Kita-ku, Kyoto 603-8555, Japan}

\author{Reiko Furusho}
\affiliation{Tsuru University, 3-8-1 Tahara, Tsuru, Yamanashi 402-0052, Japan}
\affiliation{National Astronomical Observatory of Japan, 2-21-1 Osawa, Mitaka, Tokyo 181-8588, Japan}

\author[0000-0001-6423-3768]{Daniel C. Boice}
\affiliation{Scientific Studies and Consulting, 171 Harmon Drive, San Antonio, TX 78209, USA}

\author{Tsuyoshi Terai}
\affiliation{Subaru Telescope, National Astronomical Observatory of Japan, 650 North A`ohoku Place, Hilo, HI 96720, USA }

\author[0000-0002-0717-0972]{Hirotomo Noda}
\affiliation{RISE project, National Astronomical Observatory of Japan, 2-21-1 Osawa, Mitaka, Tokyo 181-8588, Japan}

\author[0000-0001-9761-185X]{Noriyuki Namiki}
\affiliation{RISE project, National Astronomical Observatory of Japan, 2-21-1 Osawa, Mitaka, Tokyo 181-8588, Japan}

\author{Jun-ichi Watanabe}
\affiliation{Public Relation Center, National Astronomical Observatory of Japan, 2-21-1 Osawa, Mitaka, Tokyo 181-8588, Japan}



\begin{abstract}

The linear polarization degree (referred to the scattering plane, $P_{\rm r}$) as a function of the solar phase angle, $\alpha$, of solar system objects is a good diagnostic to understand the scattering properties of their surface materials. We report $P_{\rm r}$ of Phaethon over a wide range of $\alpha$ from 19$^\circ$.1 to 114$^\circ$.3 in order to better understanding properties of its surface materials. The derived phase-polarization curve shows that the maximum of $P_{\rm r}$, $P_{\rm max}$, is $>$42.4\% at $\alpha >$114$^\circ$.3, a value significantly larger than those of the moderate albedo asteroids ($P_{\rm max} \sim$9\%). The phase-polarization curve classifies Phaethon as $B$-type in the polarimetric taxonomy, being compatible with the spectral property. We compute the geometric albedo, $p_{\rm v}$, of 0.14 $\pm$ 0.04 independently by using an empirical slope-albedo relation, and the derived $p_{\rm v}$ is consistent with previous results determined from mid-infrared spectra and thermophysical modeling. We could not find a fit to the period in our polarimetric data in the range from 0 up to 7.208 $hr$ (e.g., less than twice the rotational period) and found significant differences between our $P_{\rm r}$ during the 2017 approach to the Earth and that of the 2016. These results imply that Phaethon has a region with different properties for light scattering near its orbital pole.

\end{abstract}



\section{Introduction} \label{sec:intro}

Asteroid (3200) Phaethon is an Apollo-type near-Earth asteroid with a diameter of 5.1 $\pm$ 0.2 km and a rotational period of 3.603958 $\pm$ 0.000002 $hr$ \citep{Hanus2016}. 
It has a large orbital inclination (22.2$^\circ$) and small perihelion distance (0.14 au). The surface of Phaethon has a geometric albedo, $p_{\rm v}$, of 0.122 $\pm$ 0.008 using a thermophysical model of its mid-infrared spectra \citep{Hanus2016}. 
Phaethon is thought to be a collisional family member of asteroid (2) Pallas \citep{Lemaitre_Morbidelli1994}, and they found that the typical  $p_{\rm v}$ of the members of the Pallas Collisional Family (PCF) is larger than that of other $B$-type asteroids excluding PCF \citep{Lemaitre_Morbidelli1994}. 
Phaethon is also likely the parent body of the Geminid meteor shower because of their orbital association \citep{Whipple1983, Williams_Wu1993, deLeon2010}. 
Recently, the small brightening and comet-like tails of Phaethon observed near perihelion in 2009, 2010, and 2012 were found to be caused by small dust grains produced by thermal fracture and/or desiccation cracking of surface materials and released by the solar radiation pressure \citep{Jewitt_Li2010, Li_Jewitt2013, Jewitt2013}. Because these current mass-loss events are not sufficient to explain the activity of the Geminids \citep{Li_Jewitt2013}, Phaethon probably released a large amount of dust particles in the past, possibly due to comet-like activity driven by water ice sublimation.

Because of its small perihelion distance, the surface temperature of Phaethon exceeds 1000 K \citep{Ohtsuka2009, Boice2017} and receives high solar radiation pressure near perihelion. These effects are expected to cause small grains with radius of $<$1 mm to be blown off its surface \citep{Jewitt_Li2010} and thermal metamorphism of surface materials. It is thought that its surface is covered by rocks with coarser grain size and contains hydrated minerals from various observational, experimental, and theoretical studies \citep{Licandro2007, deLeon2010, Hanus2016}. The linear polarization degree referred to the scattering plane,  $P_{\rm r}$ \citep{Zellner_Gradie1976}, as a function of the solar phase angle, $\alpha$ (i.e., Sun-object-observer angle) is a good diagnostic to understand the scattering properties of surface materials. Here we call this relation the 'phase-polarization curve'. The $P_{\rm r}$ in the large $\alpha$ region is controlled primarily by the properties of individual particles in the medium \citep{Hapke2012}. Anyway, at the lower phase angle region (at $\alpha < \sim 40^\circ$), the phase-polarization curve has been used for the polarimetric classification for asteroids \citep{Belskaya2017} and estimation of a geometric albedo, $p_{\rm v}$ \citep{Cellino2015}. Recent polarimetric results in the positive polarization branch of Phaethon during the 2016 and 2017 approach to the Earth suggest that Phaethon has an extremely high $P_{\rm r}$ compared to other solar system bodies \citep{Ito2018, Devogele2018}; however, there was no measurement of $P_{\rm r}$ of Phaethon around the inversion angle ($\alpha <30^\circ$).

\section{Observations and data reduction} \label{sec:obs}

\begin{deluxetable*}{lcccl}[t]
\tablecaption{Polarimetric results of Phaethon\label{tab:obslog}}
\tablecolumns{15}
\tablenum{1}
\tablewidth{0pt}
\tablehead{
    \colhead{UT Time in 2017}  & \colhead{$\phi$ ($^\circ$)}     & \colhead{$\alpha$ ($^\circ$)}     & \colhead{$T_{\rm exp}$ (s $\times$ sequence)}         & \colhead{Polarimetric standard stars} 
}
\startdata
Dec 9  12:16:13--17:46:49 & 203.34--200.28 &  19.31-- 19.21 & 30 $\times$ 93                    & HD 65583 (UP, GT), HD 204827 (SP) \\
Dec 10 10:58:02--16:57:46 & 189.85--185.70 &  19.12-- 19.19 & 30 $\times$ 94                    & HD 65583 (UP, GT), HD 204827 (SP) \\
Dec 11 10:46:27--16:31:51 & 172.50--167.81 &  19.81-- 20.19 & 30 $\times$ 99                    & HD 214923 (UP, GT), HD 204827 (SP) \\
Dec 12 12:20:02--16:32:54 & 151.07--147.33 &  22.28-- 22.92 & 20 $\times$  5 $+$ 30 $\times$ 72 & HD 432 (UP, GT), HD 204827 (SP) \\
Dec 13 10:15:17--15:12:28 & 131.90--127.55 &  26.43-- 27.71 & 30 $\times$ 76                    & HD 214923 (UP, GT), HD 204827 (SP) \\
Dec 14 12:11:57--15:58:44 & 110.04--107.00 &  34.46-- 35.95 & 30 $\times$ 60                    & HD 39587 (UP, GT) \\
Dec 15 09:10:53--11:01:58 & 115.53--112.38 &  43.30-- 44.56 & 20 $\times$ 44                    & HD 39587 (UP, GT), HD 19820 (SP) \\
Dec 16 09:00:58--13:17:53 &  80.00-- 77.93 &  56.68-- 59.26 & 20 $\times$ 72 $+$ 30 $\times$ 21 & HD 39587 (UP, GT), HD 19820 (SP) \\
Dec 17 09:23:30--12:34:53 &  70.52-- 69.66 &  71.50-- 73.45 & 20 $\times$ 41                    & HD 39587 (UP), HD 432 (GT), HD 19820 (SP) \\
Dec 18 08:52:14--12:03:59 &  66.01-- 65.68 &  85.18-- 86.93 & 30 $\times$ 64                    & HD 39587 (UP, GT), HD 19820 (SP) \\
Dec 19 08:57:50--11:21:10 &  64.71-- 64.70 &  97.11-- 98.22 & 30 $\times$ 44                    & HD 39587 (UP, GT), HD 19820 (SP) \\
Dec 20 08:37:34--10:28:23 &  65.29-- 65.39 & 106.54--107.19 & 60 $\times$ 20                    & HD 39587 (UP, GT), HD 19820 (SP) \\
Dec 21 08:48:10--09:48:26 &  66.83-- 66.91 & 114.03--114.30 & 120 $\times$ 7                    & HD 154345 (UP, GT), HD 19820 (SP) \\
\enddata
\tablecomments{
UT date is the mid-time of the start and final sequences. $\phi$ and $\alpha$ are the angle of the scattering plane and the solar phase angle of the observations, respectively. $T_{\rm exp}$ is exposure time of each frame in seconds and number of sequences. UP, GT, and SP in the column of polarimetric standard stars indicate unpolarized standard stars, fully linear polarized light from the Glan-Taylor prism, and strong polarized standard stars, respectively; chosen from \citet{Serkowski1974}, \citet{Schmidt1992}, and \citet{Wolff1996} for calibrating the instrumental polarization.
}
\end{deluxetable*}

The polarimetric survey of Phaethon was performed for 13 consecutive nights from UT 2017 December 9 to December 21 using the Polarimetric Imager for Comets (PICO; \citealt{Ikeda2007}) mounted on the 50-cm Telescope for Public Outreach\footnote{\url{https://www.nao.ac.jp/en/access/mitaka/facilities/50cm-telescope.html}} at Mitaka Campus of National Astronomical Observatory of Japan. We used the standard Johnson-Cousins $R_{\rm C}$-band filter (its bandpass is 483-799 nm; \citealt{Bessell2005}) for all observations of Phaethon. Due to favourable weather conditions, we obtained a high-quality data set in the range of solar phase angle from 19$^\circ$.12 through 114$^\circ$.30 (Table \ref{tab:obslog}). The ranges of heliocentric and geocentric distances of Phaethon were 1.13-0.94 au and 0.15-0.07 au, respectively. The elevation of all observations was over 28$^\circ$. Finally, 3,560 frames (890 sequences) of Phaethon were acquired during the survey. We excluded frames where stationary field stars and cosmic-rays overlapped Phaethon as judged by eye, leaving 3,248 frames (812 sequences) that were analysed. Details of PICO with the 50-cm telescope are described in Appendix A.

All selected PICO data were reduced using the standard reduction procedure for imaging observations of a point-source object (dark subtraction, flat-fielding, aperture photometry with the $APPHOT$ package) described in \citet{Ikeda2007} with custom PyRAF script that uses IRAF\footnote{IRAF is distributed by the National Optical Astronomy Observatory, which is operated by the Association of Universities for Research in Astronomy (AURA) under a cooperative agreement with the National Science Foundation.} via Python developed by our group. We also applied a moving-circular aperture to photometry of Phaethon with an elongated shape on the images taken under sidereal tracking on 2017 December 17 (developed by \citealt{Yoshida_Terai2017}). To calibrate the instrumental polarizations, we also observed unpolarized standard stars, completely polarized light obtained through a Glan-Taylor prism, and strong polarized standard stars. The degree of linear polarization, $P$, and the position angle, $\theta$, from normalized Stokes parameters, $q$ ($\equiv Q/I$) and $u$ ($\equiv U/I$), are converted by the following expressions \citep{Tinbergen1996}: $P = \sqrt{q^2 + u^2}$ and $\theta = \frac{1}{2} \arctan(u/q)$. Details of the polarization calibrations, correction of instrumental polarizations, and error estimations are described in Appendix B and \citet{Kawabata1999}.

In general, the degree of linear polarization, $P_{\rm r}$, and position angle, $\theta_{\rm r}$, in the scattering plane (the plane containing the object, the Sun, and the Earth at the time of observation) have been used to compare with other  solar system objects.  
$P_{\rm r}$ and $\theta_{\rm r}$ are expressed in $P_{\rm r} = P_{\rm cel}\cos(2\theta_{\rm cel})$ and $\theta_{\rm r} = \theta_{\rm cel} - (\phi \pm 90^\circ)$, respectively, where $\phi$ is the position angle of the scattering plane and the sign is chosen to satisfy  0$^\circ \leq (\phi \pm 90^\circ) \leq 180^\circ$ \citep{Chernova1993}. $P_{\rm cel}$ and $\theta_{\rm cel}$ are linear polarization degree and polarization position angle in celestial coordinates, respectively.
Position angles in the scattering plane at the mid-time of each sequence were calculated using JPL's HORIZONS system\footnote{\url{https://ssd.jpl.nasa.gov/horizons.cgi}}. The resultant weighted mean of $P_{\rm r}$ and $\theta_{\rm r}$ on each date was computed and is summarized in Table \ref{tab:results}.

\begin{figure}
\plotone{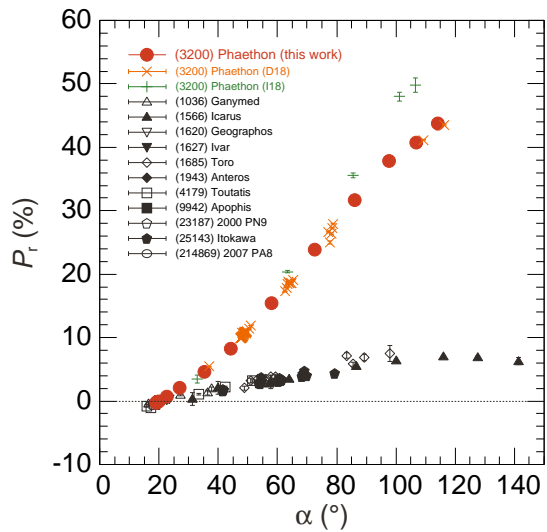}
\caption{Phase-polarization curve of Phaethon in the $R_{\rm C}$-band. Vertical and horizontal axes are the degree of linear polarization, $P_{\rm r}$, and the solar phase angle, $\alpha$, respectively. Red circles are the observed $P_{\rm r}$ of Phaethon at $\alpha$ on each date. The error bars are less than the symbol size, including both random errors (all sequences on each date and the standard deviation of polarization standard stars during the survey) and the systematic error of PICO. Orange crosses (D18) and green plus (I18) are the $P_{\rm r}$ of Phaethon taken on 2017 December \citep{Devogele2018} and during 2016 September-November \citep{Ito2018}, respectively. Black symbols are the $P_{\rm r}$ of the moderate albedo asteroids \citep{Lupishko2014, Ishiguro2017}. The horizontal black dotted line shows $P_{\rm r}$ = 0\%.
\label{fig:fig1}}
\end{figure}

\section{Results and discussion} \label{sec:res}

We report the phase-polarization curve of Phaethon over a wide range of $\alpha$ from 19$^\circ$.1 through 114$^\circ$.3 and find that $P_{\rm r}$ grows steadily through $\alpha$ of 114$^\circ$ (Figure \ref{fig:fig1}, Table \ref{tab:results}). 
The expected maximum linear polarization degree, $P_{\rm max}$, of Phaethon is $>$42.4\% (3$\sigma$ lower limit) and $\alpha_{\rm max}$, $\alpha$ at $P_{\rm max}$, is located at $>$114$^\circ$. 
This value is consistent with other polarimetric observations of Phaethon \citep{Ito2018, Devogele2018}. 
The derived $P_{\rm max}$ at $\alpha_{\rm max}$ of Phaethon are more than four times larger than those values for the moderate albedo asteroids (e.g., $P_{\rm max} <\sim$9\%; \citealt{Lupishko2014, Ishiguro2017}, at $\alpha_{\rm max} \sim100^\circ$; \citealt{Geake_Dollfus1986, Lupishko2014}), implying peculiar surface properties of Phaethon. 
To explain Phaethon's large linear polarization degree, \citet{Ito2018} pointed out the interpretations: lower $p_{\rm v}$ than the current estimations, relatively large grains, and high surface porosity.

\begin{deluxetable*}{lcccccc}
\tablecaption{Polarimetric results of Phaethon\label{tab:results}}
\tablecolumns{15}
\tablenum{2}
\tablewidth{0pt}
\tablehead{
    \colhead{UT Time in 2017}  & \colhead{JD - 2,458,000} & \colhead{$r_{\rm H}$ (au)} & \colhead{$\Delta$ (au)} & \colhead{$\alpha$ ($^\circ$)}   & \colhead{$P_{\rm r}$ (\%)}  & \colhead{$\theta_{\rm r}$ ($^\circ$)} 
}
\startdata
Dec 9  12:16:13--17:46:49 &  97.011262-- 97.249845 & 1.129--1.126 & 0.154--0.151 &  19.31-- 19.21 & -0.288  $\pm$ 0.002  &   3.8 $\pm$ 0.2 \\
Dec 10 10:58:02--16:57:46 &  97.956968-- 98.206782 & 1.115--1.111 & 0.139--0.135 &  19.12-- 19.19 & -0.217  $\pm$ 0.002  & 177.0 $\pm$ 0.3 \\
Dec 11 10:46:27--16:31:51 &  98.948924-- 99.188785 & 1.100--1.096 & 0.123--0.119 &  19.81-- 20.19 & -0.154  $\pm$ 0.002  & 136.1 $\pm$ 1.0 \\
Dec 12 12:20:02--16:32:54 & 100.013912--100.189514 & 1.083--1.080 & 0.107--0.105 &  22.28-- 22.92 &  0.652  $\pm$ 0.002  &  86.7 $\pm$ 0.1 \\
Dec 13 10:15:17--15:12:28 & 100.927280--101.133657 & 1.068--1.065 & 0.095--0.092 &  26.43-- 27.71 &  2.074  $\pm$ 0.002  &  88.8 $\pm$ 0.1 \\
Dec 14 12:11:57--15:58:44 & 102.008299--102.165787 & 1.051--1.049 & 0.082--0.081 &  34.46-- 35.95 &  4.593  $\pm$ 0.003  &  88.6 $\pm$ 0.1 \\
Dec 15 09:10:53--11:01:58 & 102.882558--102.959699 & 1.037--1.036 & 0.075--0.074 &  43.30-- 44.56 &  8.242  $\pm$ 0.004  &  88.5 $\pm$ 0.1 \\
Dec 16 09:00:58--13:17:53 & 103.875671--104.054086 & 1.021--1.018 & 0.070--0.069 &  56.68-- 59.26 & 15.453  $\pm$ 0.004  &  88.6 $\pm$ 0.1 \\
Dec 17 09:23:30--12:34:53 & 104.891319--105.024225 & 1.004--1.002 & 0.069--0.070 &  71.50-- 73.45 & 23.87   $\pm$ 0.02   &  88.3 $\pm$ 0.1 \\
Dec 18 08:52:14--12:03:59 & 105.869606--106.002766 & 0.987--0.985 & 0.074--0.075 &  85.18-- 86.93 & 31.71   $\pm$ 0.01   &  88.7 $\pm$ 0.1 \\
Dec 19 08:57:50--11:21:10 & 106.873495--106.973032 & 0.970--0.969 & 0.082--0.083 &  97.11-- 98.22 & 37.90   $\pm$ 0.03   &  88.6 $\pm$ 0.1 \\
Dec 20 08:37:34--10:28:23 & 107.859421--107.936377 & 0.953--0.952 & 0.093--0.094 & 106.54--107.19 & 40.76   $\pm$ 0.17   &  89.4 $\pm$ 0.1 \\
Dec 21 08:48:10--09:48:26 & 108.866782--108.908634 & 0.936--0.935 & 0.106--0.107 & 114.03--114.30 & 43.71   $\pm$ 0.44   &  88.4 $\pm$ 0.3 \\
\enddata
\tablecomments{
UT date and JD are the mid-time of the start and final sequences. $r_{\rm H}$ and $\Delta$  are the heliocentric and geocentric distances in au and $\alpha$ is the solar phase angle. $P_{\rm r}$ and $\theta_{\rm r}$ are the degree of linear polarization and polarization position angle referred to the scattering plane, respectively. The uncertainty in each value of $P_{\rm r}$ and $\theta_{\rm r}$ includes both random errors (all sequences on each date and standard deviation of polarimetric standard stars during the survey) and the systematic error of PICO.}
\end{deluxetable*}

\subsection{Polarimetric classification and geometric albedo}
\begin{figure*}
\gridline{\fig{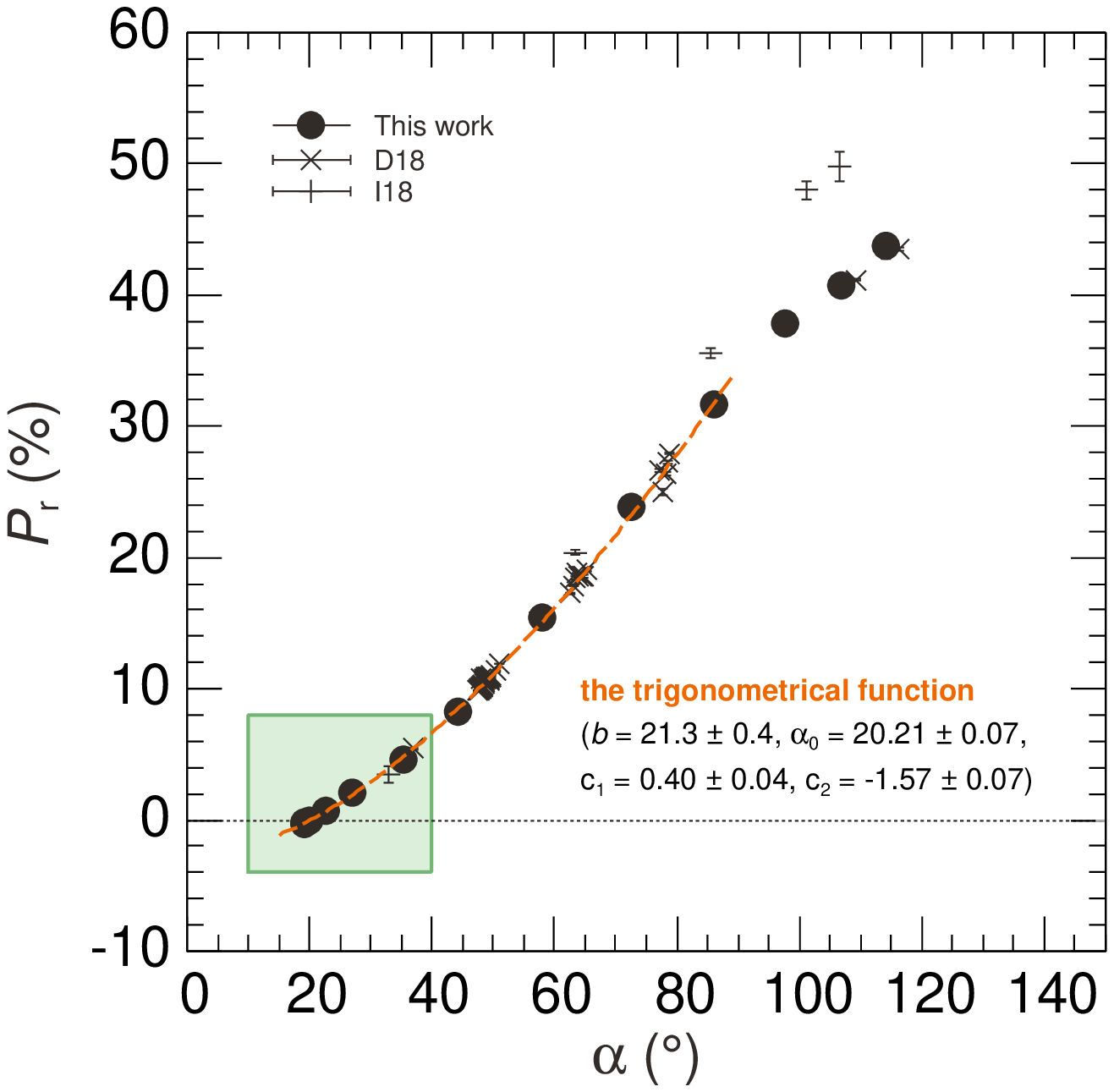}{0.45\textwidth}{(a)}
          \fig{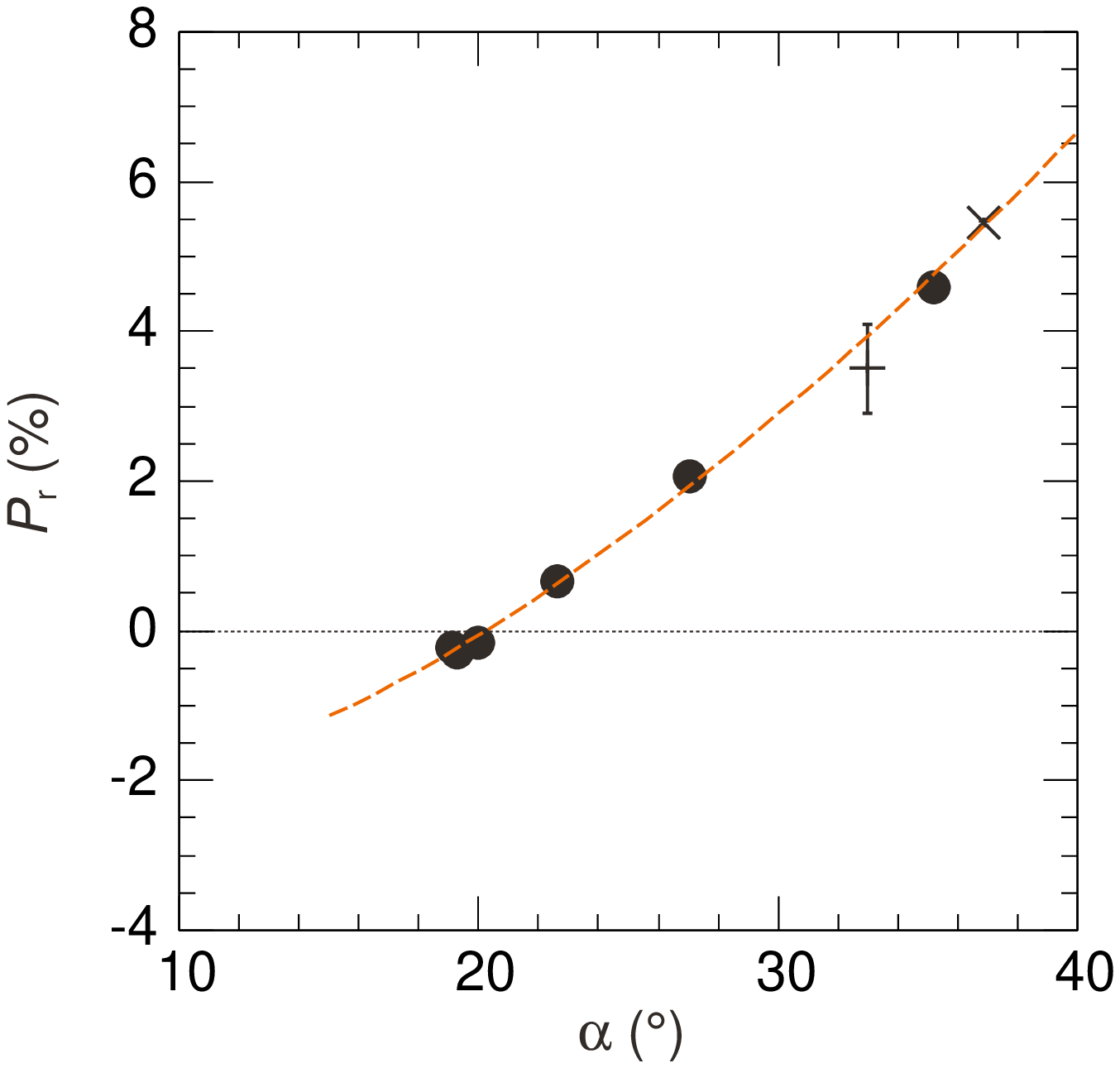}{0.45\textwidth}{(b)} }
\caption{(a) The best-fit phase-polarization curve of Phaethon in the $R_{\rm C}$-band at $\alpha<90^\circ$. (b) Expanded figure near $\alpha_{\rm inv}$ to show details (green hatch of panel (a)). Axes are identical to Figure \ref{fig:fig1}. Black circles, cross, and plus symbols are the $P_{\rm r}$ of Phaethon in this work, D18 \citep{Devogele2018}, and I18 \citep{Ito2018}, respectively. The dash (orange) lines indicate best-fit phase-polarization curve by the trigonometrical function \citep{Lumme_Muinonen1993}. Note that the fitting result is not complete in the large $\alpha$ region ($\alpha>90^\circ$) because $P_{\rm r}$ at $\alpha>90^\circ$ are not used for these fittings (see Section 3.1).
\label{fig:fig2}}
\end{figure*}

To derive an inversion angle, $\alpha_{\rm inv}$ [$^\circ$] at which $P_{\rm r}$ changes its sign, and a polarimetric slope at $\alpha_{\rm inv}$, $h$ [\% deg$^{-1}$], we computed the best-fit of the phase-polarization curve at $\alpha<90^\circ$ using the trigonometrical function \citep{Lumme_Muinonen1993} by $\chi^{2}$ minimization with the Marquardt-Levenberg algorithm \citep{Press1992} (Figure \ref{fig:fig2}). This trigonometrical function is given by $P_{\rm r}(\alpha) = b\sin^{c_{1}}(\alpha) \sin^{c_{2}}(\alpha/2) \sin(\alpha - \alpha_{\rm inv})$, where $b$, $\alpha_{\rm inv}$, $c_{1}$, and $c_{2}$ are free parameters. The trigonometrical function cannot be applied mathematically to polarimetric data when a phase-polarization curve has $\alpha>110^\circ$ \citep{Ishiguro2017}, since there is no solution of $dP(\alpha)/d\alpha$ at $\alpha>110^\circ$ with $c_{2}>0$, which $c_{2}>0$ is the original definition of this function \citep{Lumme_Muinonen1993}. As a result, we applied the trigonometrical function with $c_{2}<0$ and used only limited polarimetric data at $\alpha<90^\circ$. The derived best-fit parameters are $b$ = 21.33 $\pm$ 0.44, $\alpha_{\rm inv}$ = 20$^\circ$.21 $\pm$ 0$^\circ$.07, $c_{1}$ = 0.402 $\pm$ 0.038, and $c_{2}$ = -1.57 $\pm$ 0.07, corresponding to $\alpha_{\rm inv}$ = 20$^\circ$.21 $\pm$ 0$^\circ$.07 and $h$ = 0.174\% $\pm$ 0.053\% deg$^{-1}$. The computed minimum $P_{\rm r}$, $P_{\rm min}$, and $\alpha$ at $P_{\rm min}$, $\alpha_{\rm min}$, have large errors because there are no observations of $P_{\rm r}$ at $<$19$^\circ$ in the polarimetric data set (Figure \ref{fig:fig2} and Table 1). The derived $\alpha_{\rm inv}$ is consistent with that of Phaethon alone (18$^\circ$.8 $\pm$ 1$^\circ$.6; \citealt{Devogele2018}). We note that these fitting results are not complete in the large $\alpha$ region ($\alpha >$90$^\circ$) because of the fitting function.

We find that Phaethon is likely to be a $B$-type asteroid as well as $M$- and $K$-type asteroids by comparing our derived $\alpha_{\rm inv}$ and $h$ to the polarimetric taxonomy (Table 3 of \citealt{Belskaya2017}). Phaethon is classified as a $B$-type asteroid by the spectral classification of asteroids \citep{Bus_Binzel2002, DeMeo2009}. We also confirmed that the phase-polarization curve of Phaethon at $\alpha <40^\circ$ shows a similar trend to those of $B$- and $F$-type main-belt asteroids (Figure 4 of \citealt{Gil-Hutton_Garcia-Migani2017}). Regarding the polarimetric taxonomy, a behavior of $P_{\rm r}$ in the high-$\alpha$ region is unknown because mainly main-belt asteroids were used, which are difficult to acquire  $P_{\rm r}$ in the high-$\alpha$ region from the ground-based observatories. Moreover, the derived $\alpha_{\rm inv}$ is larger than and derived $h$ is consistent with those of Pallas ($\alpha_{\rm inv}$ = 18$^\circ$.1 $\pm$ 0$^\circ$.1 and $h$ = 0.228\% $\pm$ 0.003\% deg$^{-1}$; \citealt{Masiero2012}). \citet{Belskaya2017} claimed that asteroids with much smaller $\alpha_{\rm inv}$ have larger amount of regolith based on a relation between $P_{\rm min}$ and $\alpha_{\rm inv}$ for asteroids of variable taxonomy types overlapped with lunar bare rocks and fines reported in \citet{Geake_Dollfus1986}. Based on this relation, Phaethon should have smaller grains on its surface compared with Pallas. On the other hand, \citet{Delbo2007} pointed out that much smaller bodies have less regolith or less mature regolith. It is expected that sizes of surface materials of Phaethon are larger than those of Pallas because the diameter of Phaethon ($\sim$5 km; \citealt{Hanus2016}) is much smaller than that of Pallas ($\sim$500 km; \citealt{Carry2010}). This inconsistency implies that $\alpha_{\rm inv}$ of an asteroid reflects scattering properties of grains (e.g., complex refractive index and grain size distribution) rather than a typical size of particles or rocks on the surface. Note that the relation between $P_{\rm min}$ and $\alpha_{\rm inv}$ for asteroids may not be suitable for Phaethon because the $P_{\rm r}$ of the materials that were investigated (e.g., \citealt{Geake_Dollfus1986, Belskaya2017}) is significantly lower than that of Phaethon. 

We estimate the $p_{\rm v}$ independently by using the empirical slope-albedo relation in the standard $V$-filter given by $\log_{10}p_{\rm v} = C_{1}\log_{10}h + C_{2}$, where $h$ is the polarimetric slope at $\alpha_{\rm inv}$ [\% deg$^{-1}$]. $C_{1}$ and $C_{2}$ are constants ($C_{1}$ = -0.80 $\pm$ 0.04 and $C_{2}$ = -1.47 $\pm$ 0.04 when $p_{\rm v} \geq$ 0.08; \citealt{Cellino2015}). 
We apply this empirical relation to our polarimetric results in the $R_{\rm C}$-band because no clear difference between $P_{\rm r}$ in the $V$- and $R_{\rm C}$-bands of Phaethon have been reported \citep{Devogele2018}. 
The derived $p_{\rm v}$ is equal to 0.14 $\pm$ 0.04 using our derived value of $h$ (0.174\% $\pm$ 0.053\% deg$^{-1}$), corresponding to a moderate value among asteroids \citep{Masiero2018}.
This value is consistent with a previous measurement of Phaethon determined from mid-infrared spectra and thermophysical modeling ($p_{\rm v}$ = 0.122 $\pm$ 0.008; \citealt{Hanus2016}), and may be high compared with typical cometary values ($\sim$0.04; \citealt{Rickman2017}) claimed in \citet{Devogele2018}. This value is also consistent with $B$-type as well as $M$- and $K$-type asteroids \citep{Belskaya2017}. $P_{\rm r}$ in the low-$\alpha$ region ($\alpha$ $<$15$^\circ$) is required to derive other polarimetric parameters (e.g., $P_{\rm min}$ and $\alpha_{\rm min}$) and classify Phaethon in the polarimetric taxonomy.

To understand scattering properties of Phaethon's surface more deeply via reproducing its phase-polarization curve, we require complex refractive index as well as size distribution of dominant surface materials.  
To derive an absorption coefficient of the complex refractive index, measurement of a circular polarization degree is theoretically useful \citep{Hapke2012}. \citet{Mukai1987} demonstrated the importance of the negative as well as the positive branches of $P_{\rm r}$ to derive a typical complex refractive index of grain materials for comet 1P/Halley. Their result was based on the numerical calculations applying Mie theory and assuming a grain-size distribution obtained by the Vega mission \citep{Mazets1987}. We strongly encourage laboratory experiments to derive phase-polarization curves for various materials expected to exist on the surfaces of asteroids.

\subsection{No confirmation of periodic change of $P_{\rm r}$}

\begin{figure*}
\figurenum{3}
\gridline{\fig{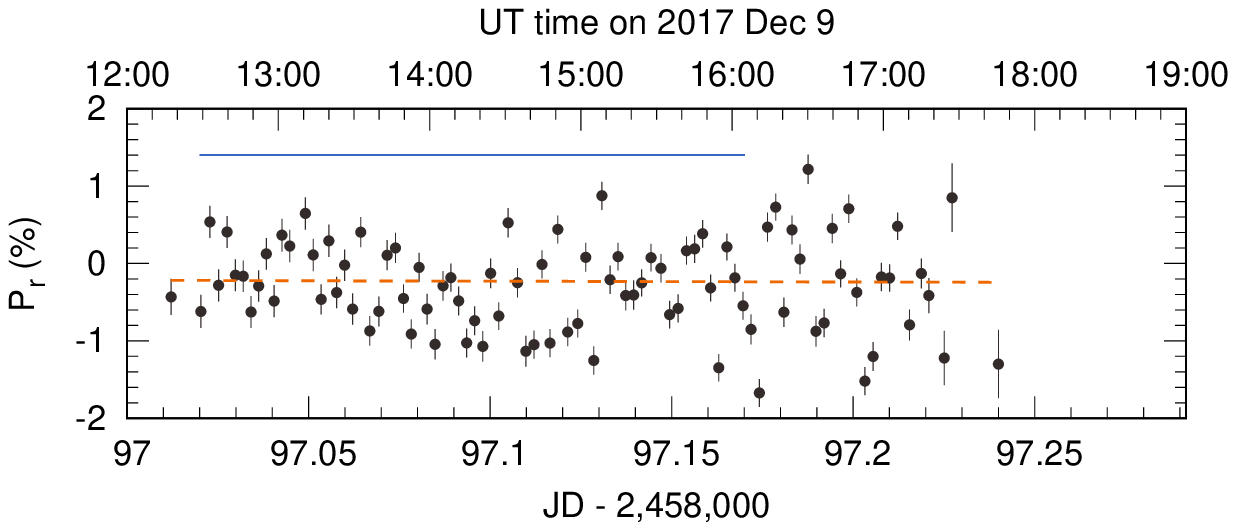}{0.49\textwidth}{(a)}
          \fig{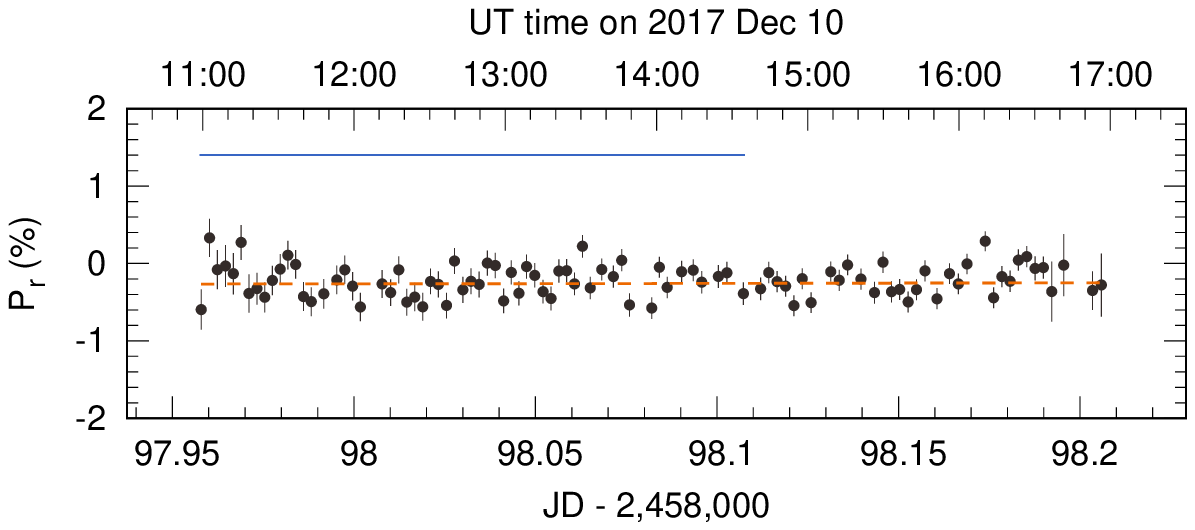}{0.49\textwidth}{(b)} }
\gridline{\fig{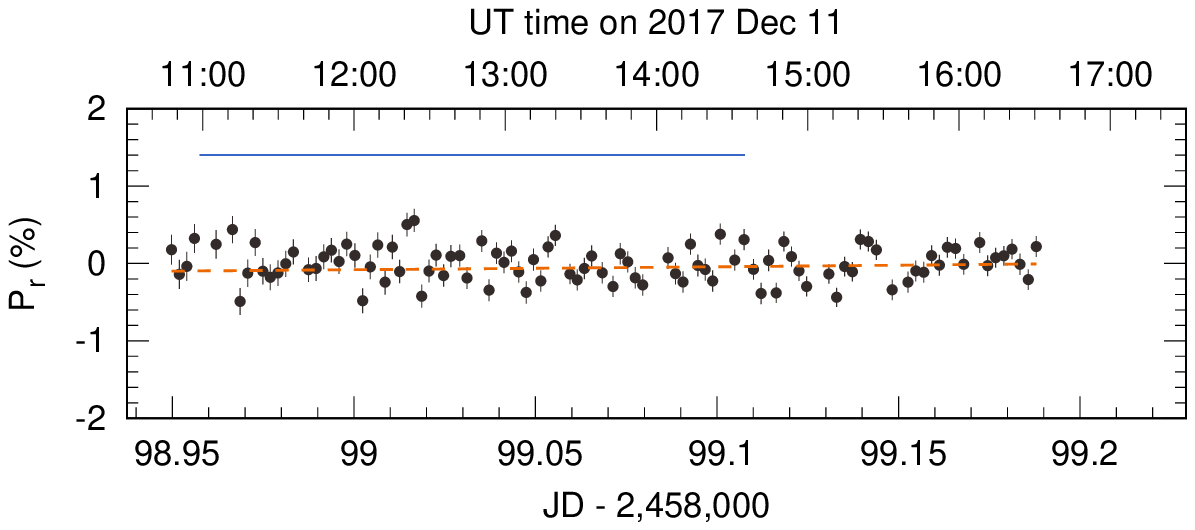}{0.49\textwidth}{(c)}
          \fig{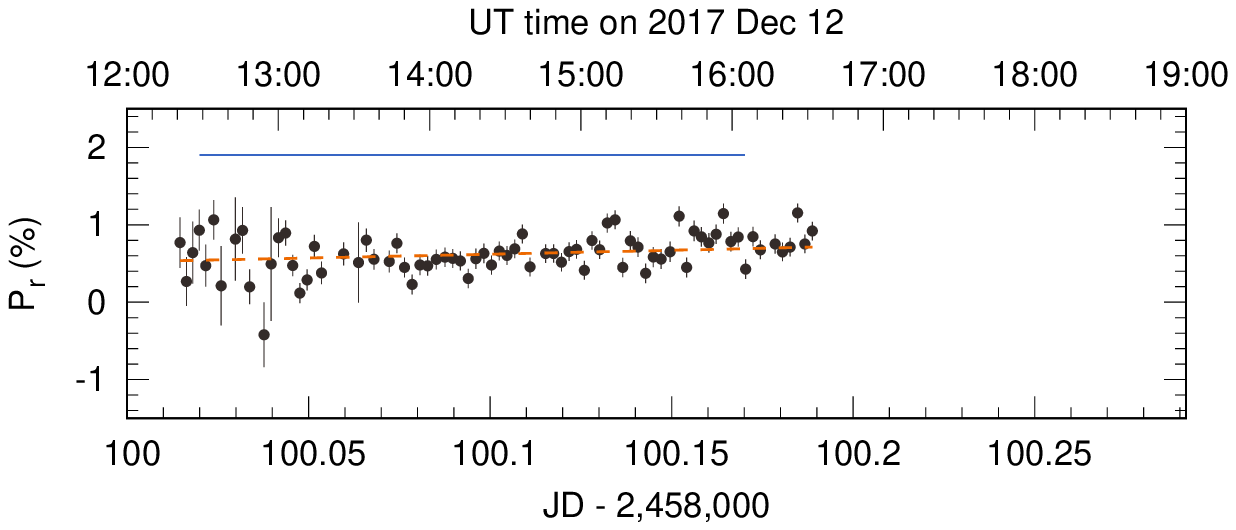}{0.49\textwidth}{(d)} }
\gridline{\fig{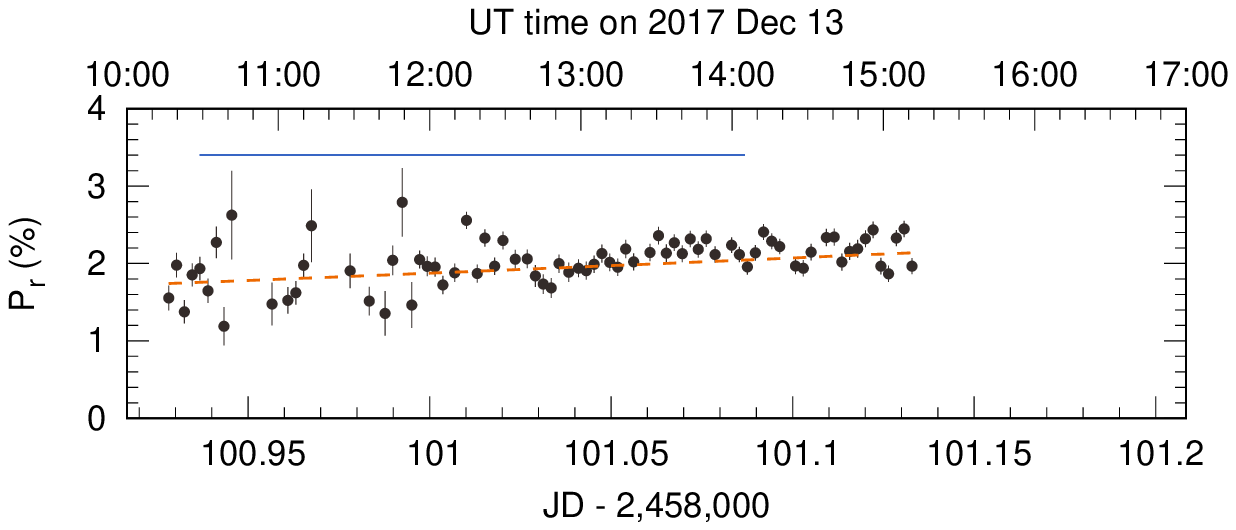}{0.49\textwidth}{(e)}
          \fig{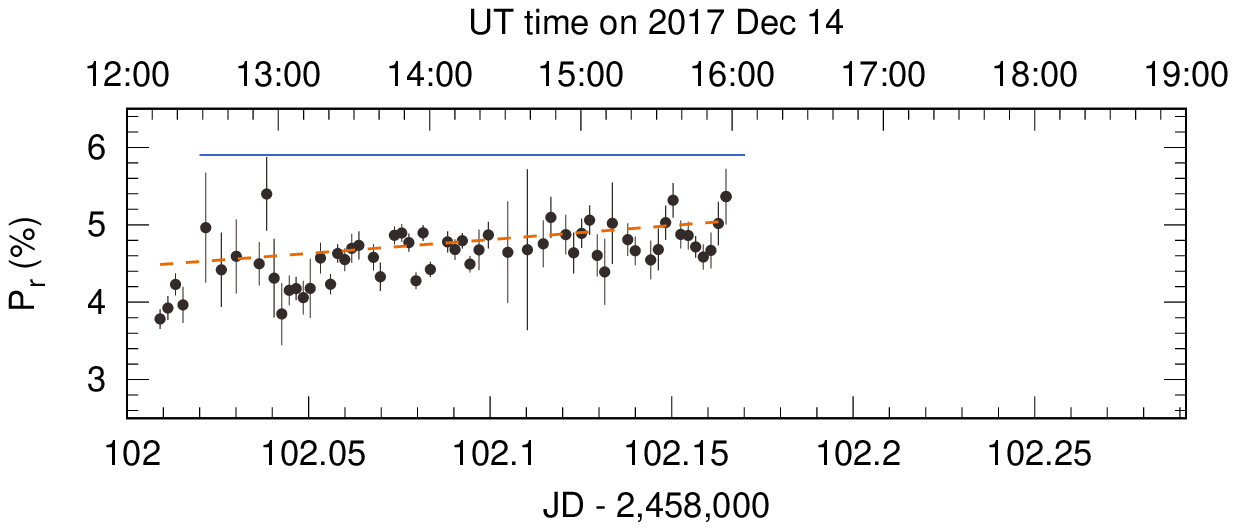}{0.49\textwidth}{(f)} }
\gridline{\fig{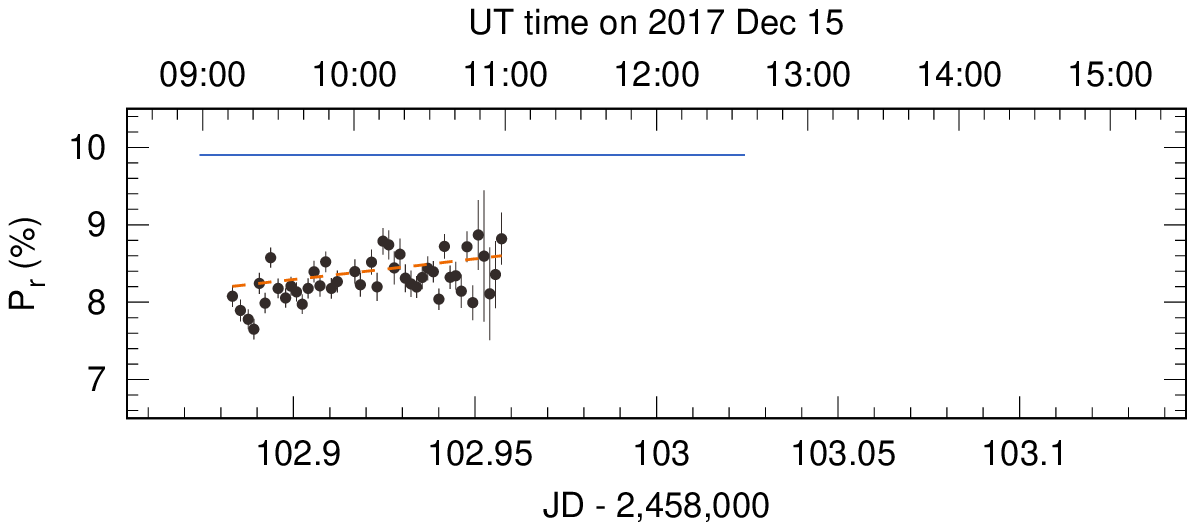}{0.49\textwidth}{(g)}
          \fig{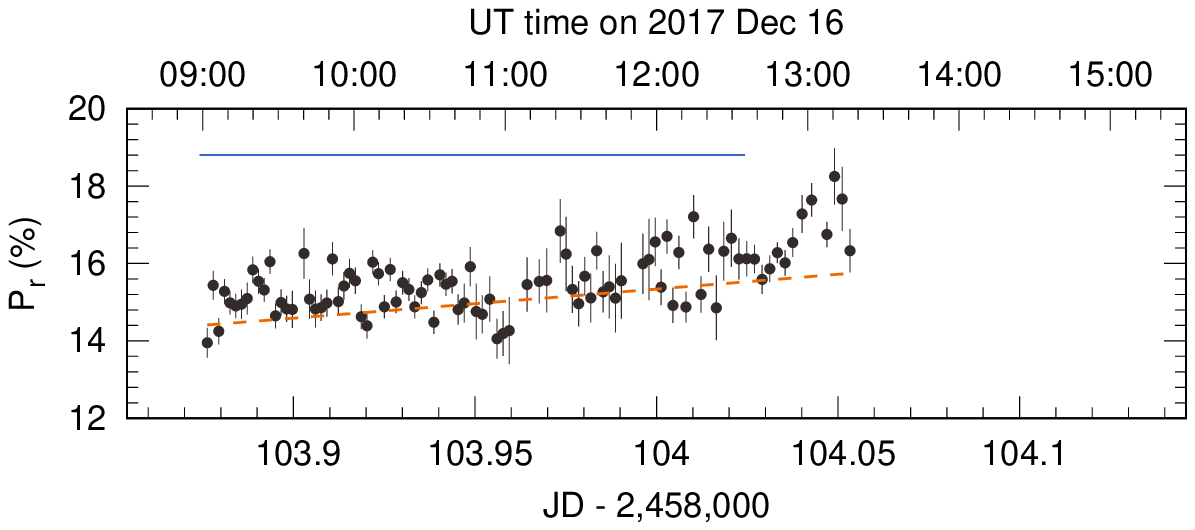}{0.49\textwidth}{(h)} }
\caption{(a-m) Time-domain $P_{\rm r}$ of Phaethon from 2017 December 9 to December 21 and (n) the phase dispersion minimization plots for  $P_{\rm r}$ of Phaethon from 2017 December 9 to 18. For panels (a) to (m), Black circles indicate $P_{\rm r}$ of Phaethon in each sequence in this work. The uncertainty of each plot includes both random and systematic errors described in Appendix B. The length of the upper horizontal (blue) bars indicate the rotational period of 3.604 $hr$. Orange dashed-lines in the panels of 2017 December 9-18 indicate phase-polarization fits using the trigonometrical function \citep{Lumme_Muinonen1993} in the range of $0^\circ < \alpha < 90^\circ$ (see section 3.1). The values and errors of each plot are listed in Table 3. For panel (n), vertical and horizontal axis are the dispersion of PDM, $\Theta$, and the orbital period in hours, respectively. The horizontal gray dotted-line indicates $\Theta$ = 1.0. We cannot find any best-fit period in the range from 0 up to 7.208 $hr$ (e.g., less than twice the rotational period), since there is no orbital period with small $\Theta$ lower than 1.0.
\label{fig:fig3}
}
\end{figure*}

\begin{figure*}
\figurenum{3}
\gridline{\fig{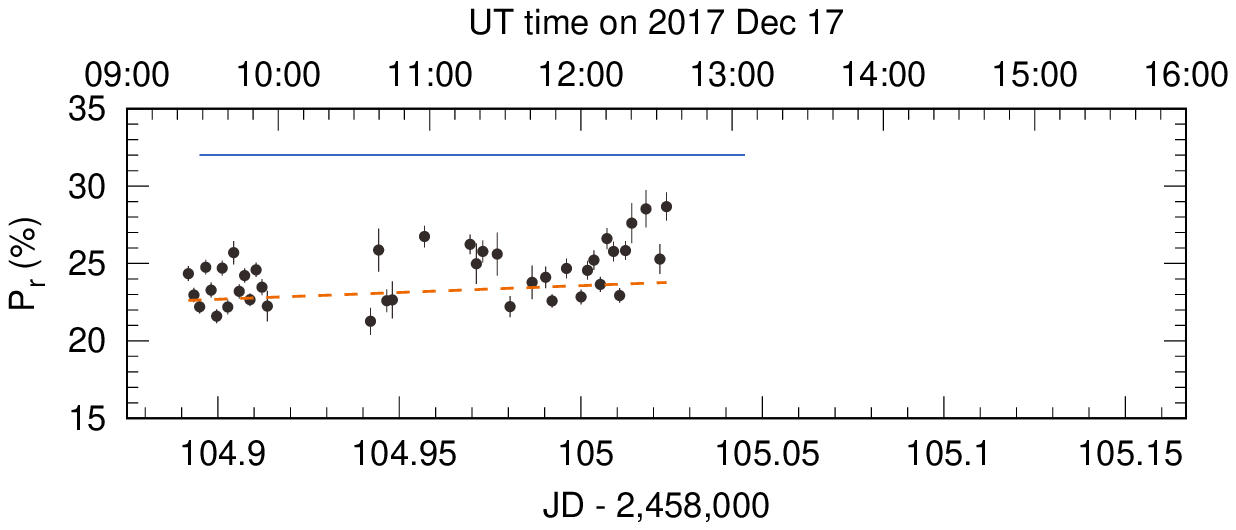}{0.49\textwidth}{(i)}
          \fig{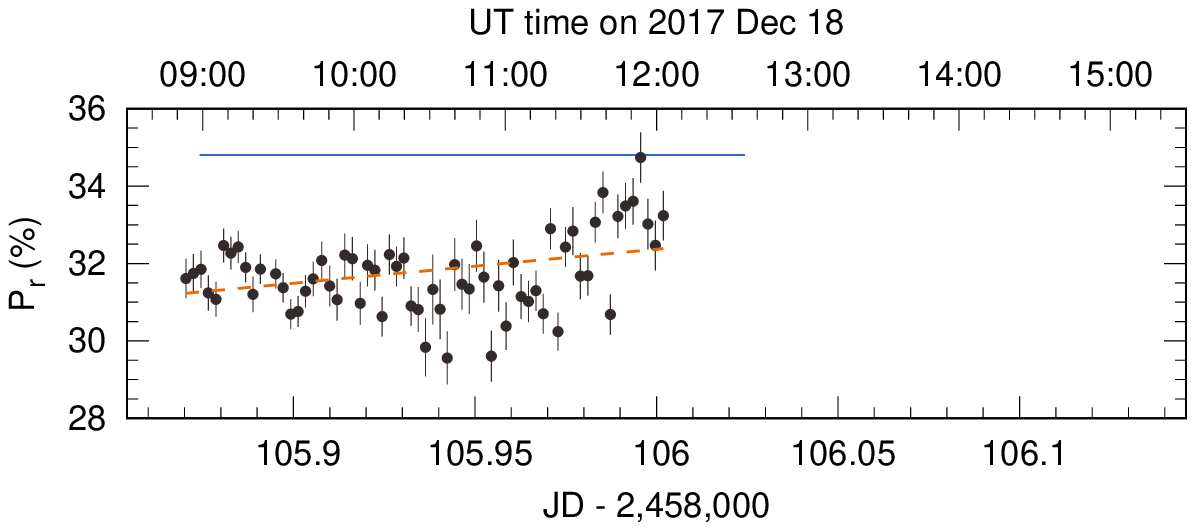}{0.49\textwidth}{(j)} }
\gridline{\fig{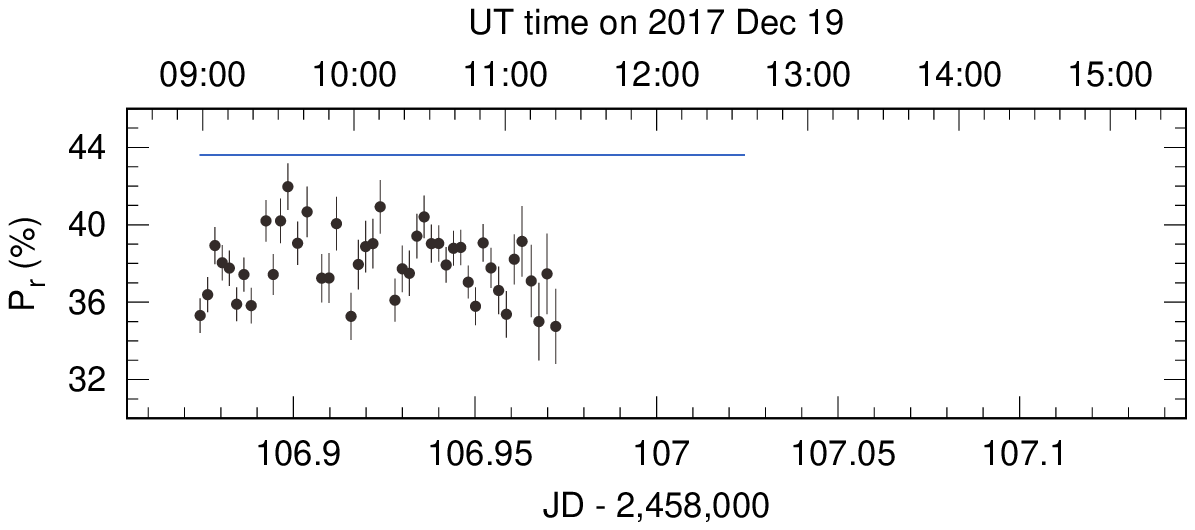}{0.49\textwidth}{(k)}
          \fig{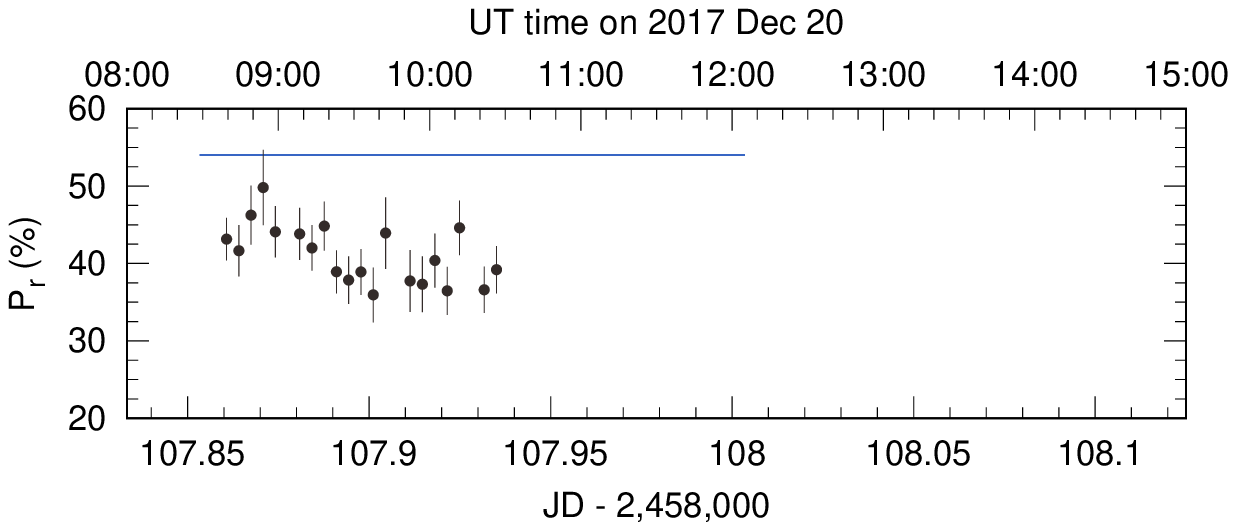}{0.49\textwidth}{(l)} }
\gridline{\fig{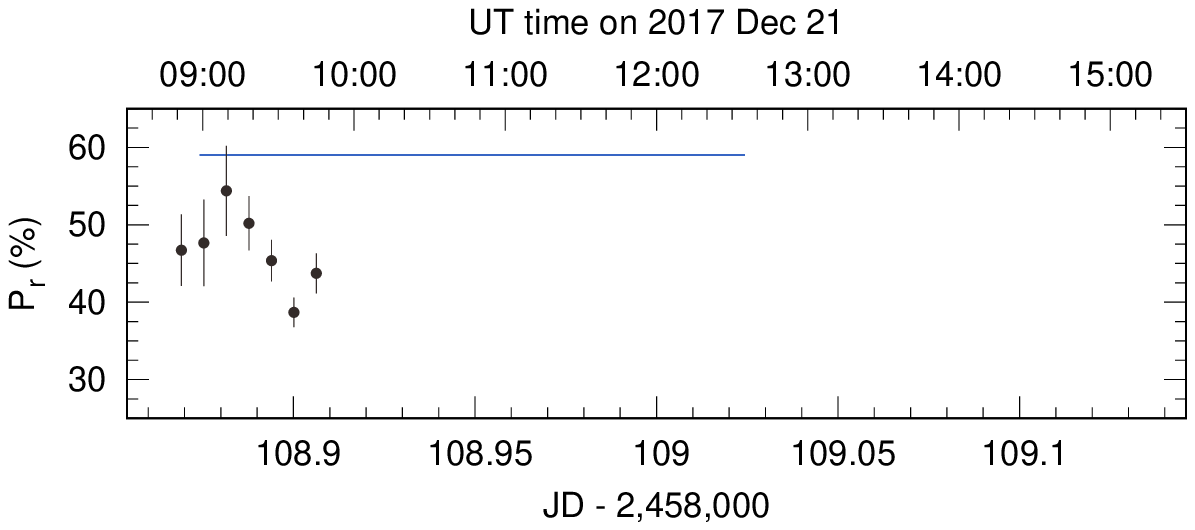}{0.49\textwidth}{(m)}
          \fig{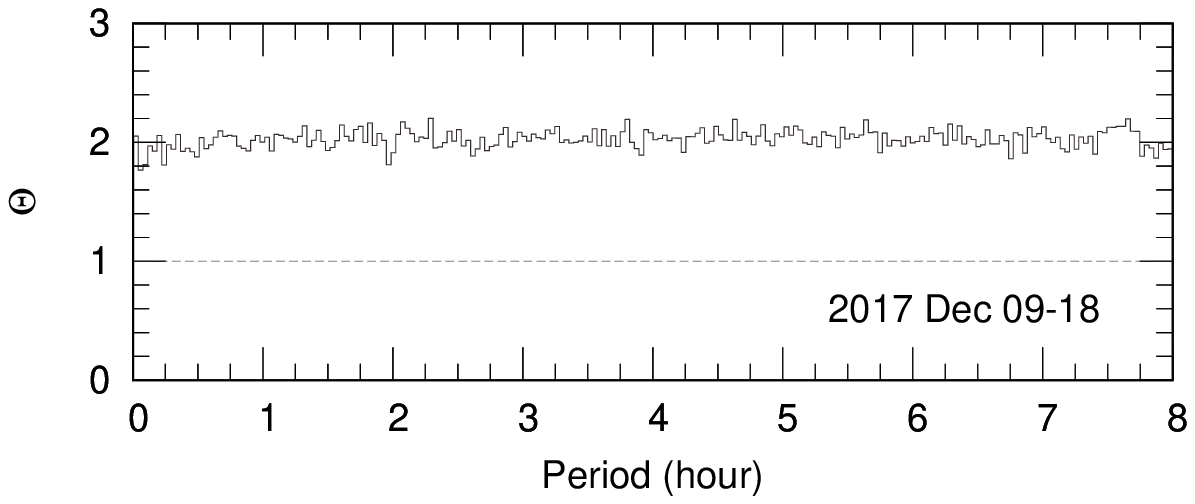}{0.49\textwidth}{(n)} }
\caption{{\it continued.}\label{fig:fig3}
}
\end{figure*}

\citet{Degewij1979} reported a variation in polarization degree in the $B$-band correlated with the lightcurve for (4) Vesta. The polarization variation was interpreted as albedo inhomogeneities of its surface materials \citep{Degewij1979}. Time-resolved polarimetry is an effective method to investigate albedo heterogeneity on asteroids. 
Panels (a)-(m) of Figure \ref{fig:fig3} show time-domain $P_{\rm r}$ of Phaethon during our polarimetric survey. 
We employed the phase dispersion minimization (PDM) method \citep{Stellingwerf1978} to search for periodicity in our polarimetric data, applying the 'cyclocode' software\footnote{\url{http://www.toybox.rgr.jp/mp366/lightcurve/cyclocode/cyclocode.html} developed by \citet{Dermawan2004}}. 
The best-fit period should have a very small normalized dispersion, $\Theta$, compared with the unphased data, and thus $\Theta \ll$1 indicates that a good fit has been found. To apply PDM fitting, we extract variable components from all $P_{\rm r}$ values until 2017 December 18 at $\alpha <90^\circ$, with the best-fit phase-polarization curve derived by the trigonometrical function (see section 3.1 and Figure \ref{fig:fig3} (a-j)). 
Panel (n) of Figure \ref{fig:fig3} shows a PDM plot for variable components of $P_{\rm r}$ of Phaethon from 2017 December 9 to 18. We cannot find any good fit to the period in our polarimetric data (e.g., individual or combined dates from 2017 December 9 to 18), although \citet{Borisov2018} found a variation of $P_{\rm r}$ with rotation on 2017 December 15. Note that our polarimetric data on December 15 does not cover a complete period because of weather conditions. Using the pole orientation of ($\lambda_{\rm pole}$, $\beta_{\rm pole}$) = (319$^\circ$, -39$^\circ$) with a 5$^\circ$ uncertainty \citep{Hanus2016}, a surface region as seen from the Earth crossed from edge-on (perpendicular to pole direction) to the near north-pole direction during our survey by considering the positional relation between Earth and Phaethon at the time of these observations. Before 2017 December 18, we observed the edge-on direction (e.g., a different part of the surface at the rotational phase). This result suggests that surface materials on most regions of Phaethon have similar scattering properties (e.g., $p_{\rm v}$).

Focusing on Phaethon's polarimetric results in the $R_{\rm C}$-band, Figure \ref{fig:fig1} shows that our phase-polarization curve is significantly different in the high-$\alpha$ region ($\alpha > \sim 60^\circ$) with that during 2016 \citep{Ito2018}, although it is in agreement with that during 2016 in the low-$\alpha$ region \citep{Ito2018} and during 2017 December \citep{Devogele2018}. Polarimetry in 2016 \citep{Ito2018} observed only the edge-on direction (e.g., $\sim$-80$^\circ$ inclination of the north-pole direction seen from the Earth).  In this case, materials causing lower polarization in the high-$\alpha$ regions by scattering are distributed near its rotational pole, and most surface regions (except for near the rotational pole region) have similar scattering properties. 
The difference of scattering properties on Phaethon's surface may reflect formation conditions in the early solar nebula and/or surface alternation after formation by the solar heating near perihelion. 
It is expected that not only determination of pole orientation but also the variation of scattering properties of materials with location on Phaethon's surface will be elucidated by detailed observations of the close flyby of Phaethon by the DESTINY$^{+}$ mission \citep{Sarli2018}.

\acknowledgments

We are grateful to the staff of the Public Relation Center of the National Astronomical Observatory of Japan for their support during our observations. The authors sincerely thank Dr. Y. Ikeda, Prof. H. Kawakita, and Prof. M. Ishiguro for their valuable advice and comments. This research was supported by Grant-in-Aid for Japan Society for the Promotion of Science (JSPS) Fellows Grant No. 15J10864 (YS), JSPS KAKENHI Grant No. 17H06459 (NN), and National Science Foundation Planetary Astronomy Program (USA) Grant No. 0908529 (DCB). The transmittance of the standard Johnson-Cousins $R_{\rm C}$-band filter made by TOPTEC (Czech Republic) was measured by using a UV-IR absorption spectrophotometer (Shimadzu, UV-3100PC) at the Advanced Technology Center of National Astronomical Observatory of Japan.

\appendix

\section{Polarimetric Imager for Comets: PICO}

PICO is a double-beam polarimetric imager with a calcite Wollaston prism and rotatable half-wave plate \citep{Ikeda2007}. 
It obtains a pair of polarized images simultaneously that are perpendicularly polarized rays (ordinary- and extraordinary-rays) split by the Wollaston prism. 
A commercial SBIG STL-1001E CCD camera was used as the detector. 
The array was cooled to -25$^\circ$C within $\pm$0.1$^\circ$C by a two-stage thermoelectric system. Before our polarimetric survey of Phaethon on December 2017, we installed the standard Johnson-Cousins $R_{\rm C}$ filter made by TOPTEC (Czech Republic). When PICO is mounted on the 50-cm Telescope for Public Outreach (F/12.06), the typical pixel scale is 0''.82 per pixel and the effective field-of-view of each polarized ray is $\sim$6'.2 $\times$ 12'.9 on the sky. 
To correct both a difference in transmittance of lenses used in PICO between ordinary- and extraordinary-rays and the time-dependent sky conditions, we obtained images at four different position angles of the half-wave plate \citep{Kawabata1999}. One set of exposures at these four different angles (0$^\circ$, 45$^\circ$, 22$^\circ$.5, 67$^\circ$.5) is called a sequence.

\section{Derivation of linear polarization degree}

After extracting counts for the eight independent images (pairs of perpendicularly polarized images at four position angles of the half-wave plate), we calculate normalized Stokes parameters, $q$ ($\equiv Q$/$I$) and $u$ ($\equiv U$/$I$), of the $i$-th sequence in instrumental coordinates as follows:

\begin{equation}
q_{\rm inst} \equiv \frac{Q(i)}{I(i)} = \frac{1-a_{1}(i)}{1+a_{1}(i)}
\end{equation}
and 

\begin{equation}
u_{\rm inst} \equiv \frac{U(i)}{I(i)} = \frac{1-a_{2}(i)}{1+a_{2}(i)}
\end{equation}
with 

\begin{equation}
a_{1}(i) = \sqrt{ \frac{ I_{\rm e,0^\circ}(i) / I_{\rm o,0^\circ}(i) }{ I_{\rm e,45^\circ}(i) / I_{\rm o,45^\circ}(i) } }
\end{equation}
and 

\begin{equation}
a_{2}(i) = \sqrt{ \frac{ I_{\rm e,22.5^\circ}(i) / I_{\rm o,22.5^\circ}(i) }{ I_{\rm e,67.5^\circ}(i) / I_{\rm o,67.5^\circ}(i) } } ,
\end{equation}
where $I_{\rm o, \Psi}$ and $I_{\rm e, \Psi}$ are the ordinary and extraordinary intensities at the angle of the half-wave plate, $\Psi$, of the $i$-th sequence \citep{Kawabata1999}. 
   In order to calibrate the instrumental polarizations (offset from the zero-point of $q$ and $u$, instrumental depolarization, and offset in the position angle between the celestial and instrumental coordinates), we also observed unpolarized standard stars, completely polarized light obtained through a Glan-Taylor prism, and strong polarized standard stars. After calibrating the instrumental polarizations, the derived $q$ and $u$ with systematic mean errors are in celestial coordinates ($q_{\rm cel} \pm \sigma_{q_{\rm cel}}$ and $u_{\rm cel} \pm \sigma_{u_{\rm cel}}$). We obtain multiple sequences of Phaethon on each date and calculated the weighted mean of $q_{\rm cel}$ and $u_{\rm cel}$ on each date as given by
   
\begin{equation}
\overline{ q_{\rm cel} } = \frac{\sum_{i=1}^{n} (q_{\rm cel}(i) / \sigma_{q_{\rm cel}}^2(i) )}{\sum_{i=1}^{n} (1 / \sigma_{q_{\rm cel}}^2(i) )}
\end{equation}
and 

\begin{equation}
\overline{ u_{\rm cel} } = \frac{\sum_{i=1}^{n} (u_{\rm cel}(i) / \sigma_{u_{\rm cel}}^2(i) )}{\sum_{i=1}^{n} (1 / \sigma_{u_{\rm cel}}^2(i) )}
\end{equation}
with an error of 

\begin{equation}
\overline{ \sigma_{q_{\rm cel}} } = \sqrt{ \frac{1}{\sum_{i=1}^{n} (1 / \sigma_{q_{\rm cel}}^{2}(i) ) } }
\end{equation}
and 

\begin{equation}
\overline{ \sigma_{u_{\rm cel}} } = \sqrt{ \frac{1}{\sum_{i=1}^{n} (1 / \sigma_{u_{\rm cel}}^{2}(i) ) } } ,
\end{equation}
respectively. Here the degree of linear polarization, $P$, and the position angle, $\theta$, from normalized Stokes parameters ($q$ and $u$) are converted by the following expressions \citep{Tinbergen1996}: $P = \sqrt{q^2 + u^2}$ and $\theta = \frac{1}{2} {\rm atan}(u/q)$. The systematic error of PICO was estimated to be $\delta P_{\rm sys}$ = ($P$/85)\% over the entire field-of-view \citep{Ikeda2007}. More details of the polarization calibrations, correction of instrumental polarizations, and error estimations are described in \citet{Kawabata1999}.

\section{Polarimetric Results of Phaethon in the $R_{\rm C}$-band}

Table 3 is the time-domain summary of the polarimetric results of Phaethon in the $R_{\rm C}$-band (812 sequences in total) taken by PICO polarimeter mounted on the 50-cm Telescope for Public Outreach at Mitaka Campus of National Astronomical Observatory of Japan.

\startlongtable
\begin{deluxetable}{lcccc}
\tablecaption{Observational results acquired with PICO\label{tab:alldata}}
\tablecolumns{15}
\tablenum{3}
\tablewidth{0pt}
\tablehead{
    \colhead{UT Date}  & \colhead{JD - 2,458,000} & \colhead{$\alpha$ ($^\circ$)} & \colhead{$P_{\rm r}$ (\%)} & \colhead{$\theta_{\rm r}$ ($^\circ$)}   
}
\startdata
2017 Dec 9	&	97.0122 	&	19.313 	&	-0.43 	$\pm$	0.23 	&	155	$\pm$	10	\\
2017 Dec 9	&	97.0203 	&	19.309 	&	-0.62 	$\pm$	0.21 	&	167	$\pm$	9	\\
2017 Dec 9	&	97.0228 	&	19.308 	&	0.54 	$\pm$	0.21 	&	65	$\pm$	7	\\
2017 Dec 9	&	97.0252 	&	19.307 	&	-0.28 	$\pm$	0.21 	&	30	$\pm$	11	\\
2017 Dec 9	&	97.0275 	&	19.306 	&	0.41 	$\pm$	0.21 	&	79	$\pm$	13	\\
2017 Dec 9	&	97.0298 	&	19.304 	&	-0.15 	$\pm$	0.21 	&	157	$\pm$	27	\\
2017 Dec 9	&	97.0320 	&	19.303 	&	-0.16 	$\pm$	0.20 	&	142	$\pm$	8	\\
2017 Dec 9	&	97.0341 	&	19.302 	&	-0.63 	$\pm$	0.21 	&	30.3	$\pm$	4.6	\\
\enddata
\tablecomments{
JD is Julian date at the middle time of each sequence. $\alpha$ is the solar phase angle of Phaethon. $P_{\rm r}$ and $\theta_{\rm r}$ are the degree of linear polarization and position angle in the scattering plane, respectively. 
}
\end{deluxetable}

\end{document}